\documentclass[aps,prb,preprint,groupedaddress,showpacs,amsmath,amssymb]{revtex4-2}

\usepackage{graphicx}
\usepackage{dcolumn}
\usepackage{bm}
\usepackage{hyperref}
\usepackage{xcolor}

\begin{document}

\title{Hydrostatic pressure effect on Co-based honeycomb magnet BaCo$_2$(AsO$_4$)$_2$ }

\author{Shuyuan Huyan,$^{1}$ Juan Schmidt,$^{1,2}$ Elena Gati,$^3$ Ruidan Zhong,$^4$  Robert J. Cava,$^4$ Paul C. Canfield,$^{1,2}$ and Sergey L. Bud'ko$^{1,2}$}

\affiliation{$^{1}$Ames Laboratory, US DOE, Iowa State University, Ames, Iowa 50011, USA}
\affiliation{$^{2}$Department of Physics and Astronomy, Iowa State University, Ames, Iowa 50011, USA}
\affiliation{$^{3}$Max Planck Institute for Chemical Physics of Solids, 01187 Dresden, Germany}
\affiliation{$^{4}$Department of Chemistry, Princeton University, Princeton, NJ 08544, USA}

\date{\today}

\begin{abstract}
The honeycomb antiferromagnet BaCo$_2$(AsO$_4$)$_2$, in which small in-plane magnetic fields ($H1\,\approx\,$0.26\,T and $H2\,\approx\,$0.52\,T at $T\,=\,$1.8\,K$\,<T_N\,\approx\,5.4$\,K) induce two magnetic phase transitions, has attracted attention as a possible candidate material for the realization of Kitaev physics based on the 3$d$ element Co$^{2+}$. Here, we report on the change of the transition temperature $T_N$ and the critical fields $H1$ and $H2$ of BaCo$_2$(AsO$_4$)$_2$ with hydrostatic pressure up to $\sim$\,20\,kbar, as determined from magnetization and specific heat measurements. Within this pressure range, a marginal increase of the magnetic ordering temperature is observed. At the same time, the critical fields are changed significantly  (up to $\sim\,$25\,-35\,\%). Specifically, we find that $H1$ is increased with hydrostatic pressure, i.e., the antiferromagnetic state is stabilized with hydrostatic pressure, whereas $H2$, which was previously associated with a transition into a proposed Kitaev spin liquid state, decreases with increasing pressure. These results put constraints on the magnetic models that are used to describe the low-temperature magnetic properties of BaCo$_2$(AsO$_4$)$_2$.
\end{abstract}


\maketitle

\section{Introduction}

The search for quantum spin liquids, a particularly exotic state of matter, has seen a boost over the last 15 years by the postulation of (i) the analytically exactly solvable Kitaev model \cite{kit06a} that hosts a spin-liquid ground state, as well as (ii) the Jackeli-Khaliullin approach to design Kitaev interactions in real materials \cite{jac09a}. The latter initially involved spin-orbit-coupled $d^5$ ions in an octahedral environment in Mott insulators, in which the combination of crystal field effects and spin-orbit coupling coupling generates an effective $j_{\textrm{eff}}\,=\,$1/2 moment. These moments are then coupled via bond-dependent ferromagnetic Ising interactions that are the central ingredient of the Kitaev model. Following  these discoveries, the research concentrated on material realizations based on heavy, 4$d$ and 5$d$, transition metals \cite{win17a} due to their inherently large spin-orbit coupling. Prominent examples include Na$_2$IrO$_3$ \cite{sin10a}, $\alpha$-Li$_2$IrO$_3$ \cite{sin12a} and $\alpha$-RuCl$_3$ \cite{plu14a}. All the listed materials show antiferromagnetic order at finite temperatures \cite{cho12a,joh15a}, which is thought to obscure the observation of the putative Kitaev spin liquid. The emergence of magnetic order is believed to originate from the presence of sizable non-Kitaev interactions \cite{win16a,rau14a}, such as nearest-neighbor and third nearest-neighbor Heisenberg couplings as well as off-diagonal couplings. The existence of these couplings is generally attributed to the spatial extent of the 4$d$ and 5$d$ orbitals which promotes longer-range interactions. Nonetheless, theoretical and experimental efforts have suggested that appropriate tuning via magnetic fields \cite{yad16a,liu18b,bae17a,jan18a,zhe17a} or pressure \cite{yad18a,kai21a} can suppress non-Kitaev interactions opening the possibility to potentially discover the enigmatic Kitaev quantum spin liquid in a real material.

\begin{figure}
\begin{center}
\includegraphics[angle=0,width=120mm]{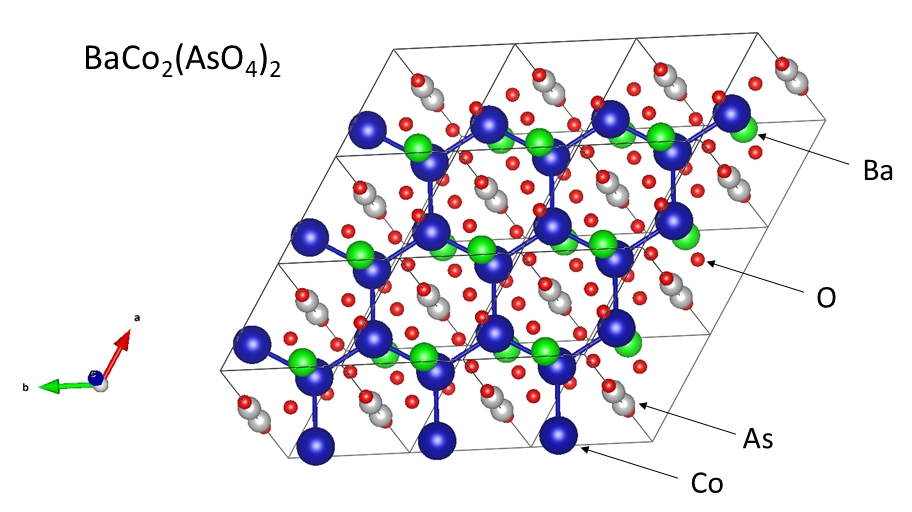}
\end{center}
\caption{(color online) A sketch of the crystal structure of BaCo$_2$(AsO$_4$)$_2$. The colors green, blue, grey, and red correspond to Ba, Co, As, and O atoms respectively. The honeycomb arrangement of the Co atoms is highlighted. Note that only half of the unit cell is shown along the $c$ axis. The structure was drawn using VESTA. \cite{mom11a} } \label{F0}
\end{figure}

In order to identify materials that might be closer to realization of the Kitaev model without further perturbation, new material design strategies are coming into the focus. A specific one involves 3$d$ transition-metal based compounds in the high-spin d$^7$ configuration, such as Co$^{2+}$ \cite{liu18a,san18a,liu20a,liu21a,kim22a}. Here, the idea is that, despite smaller spin-orbit coupling as in the 4$d$ and 5$d$ cases, the symmetry-allowed existence of ferromagnetic Heisenberg exchange of the $e_g$ electrons in the $d^7$ configuration might largely compensate for the antiferromagnetic Heisenberg interaction from the $t_{2g}$ orbitals, thus strengthening the Kitaev interaction with respect to non-Kitaev type of interactions. In this context, honeycomb BaCo$_2$(AsO$_4$)$_2$ (see Fig. \ref{F0} for the crystallographic structure) was suggested to be a particularly interesting reference compound \cite{reg77a,reg78a,reg79a,reg83a,reg90a,reg18a}. This material manifests an antiferromagnetic ground state at ambient pressure below $T_N\,\approx\,$5.4\,K. Upon small in-plane magnetic fields (note that no discernable in-plane anisotropy was reported), two consecutive magnetic transitions at $H1\,\approx\,$0.26\,T and $H2\,\approx\,$0.52\,T at $T\,=\,$1.8\,K were observed, with the former corresponding to a transition into a ferrimagnetic state and the latter potentially giving rise to a moment bearing state that has no long-range order, but is still different from a trivial, fully saturated metamagnetic state, that might be in this case the realization of the Kitaev spin liquid \cite{zho20a,zha21a}. This notion is motivated by the analogy of the temperature-field phase diagram to the one of $\alpha$-RuCl$_3$ where in-plane fields of $\sim 80$ kOe are needed to suppress magnetic order and where signatures of the Kitaev spin liquid might have been revealed \cite{yok21a}. The much smaller critical field of BaCo$_2$(AsO$_4$)$_2$ was interpreted as an experimental demonstration that non-Kitaev interactions are indeed weaker in this 3$d$ compound compared to the heavy transition-metal based compounds. Recently, time-domain terahertz spectroscopy measurements \cite{zha21a} were argued to be consistent with the behavior expected for the Kitaev model.

The proximity of BaCo$_2$(AsO$_4$)$_2$ to a putative Kitaev spin liquid state motivates to explore the possibility to further fine-tune its interactions. Applied magnetic field is only one of the external parameters that can affect the interactions. Other common experimental "knobs" include chemical and/or physical pressure, strain and either controlled or accidental disorder (and their combinations). Of those, hydrostatic pressure is considered to be both clean (not inherently inducing disorder) and not inherently symmetry breaking  parameter.  Experimentally, it allows for accommodation of different measurements, even for insulating samples. \cite{ere96a,hol97a}

In the context of pressure tuning, it should be noted that theoretical \cite{yad18a,kai21a} and experimental work \cite{sim18a,maj18a,bie18a,bas18a,wan18a,vei19a,bie20a} on the 4$d$ and 5$d$ Kitaev materials have already established that the geometry-sensitive exchange couplings are very susceptible to pressure tuning. For example, it has been pointed out that in $\alpha$-RuCl$_3$ uniaxial pressure along the $c$ axis can strengthen the Kitaev interactions while simultaneously weakening other anisotropic couplings \cite{kai21a}. However, experimentally, structural changes at low pressures \cite{maj18a,bie18a,bas18a,wan18a,vei19a}, which cause a breakdown of the Heisenberg-Kitaev picture, intervene in $\alpha$-RuCl$_3$ as well as other Kitaev compounds before a Kitaev spin liquid can be stabilized. In this regard, if the Co$^{2+}$ compounds were less susceptible to structural distortions under pressure, then they would not only offer the chance to fine-tune the balance between Kitaev and non-Kitaev interactions, but would also allow to improve the understanding of magnetoelastic coupling \cite{met20a,yem20a} in spin-orbit-coupled magnets in general.

In this work, we study the effect of hydrostatic pressure up to $\sim\,$20\,kbar on the temperature-magnetic field phase diagram of BaCo$_2$(AsO$_4$)$_2$ experimentally. We find that pressure only moderately affects the ordering temperature $T_N$, whereas it significantly changes the in-plane critical fields $H1$ and $H2$. In particular, $H1$ is increased with hydrostatic pressure, whereas $H2$ is decreased. This implies that the antiferromagnetic state as well as the proposed Kitaev spin liquid are both stabilized at the expense of the intermediate-field ferrimagnetic state by increasing applied hydrostatic pressure, which therefore puts constraints on effective models that are used to understand the magnetic behavior of BaCo$_2$(AsO$_4$)$_2$.

\section{Experimental details}

Single crystals of BaCo$_2$(AsO$_4$)$_2$ were grown using NaCl flux and pre-reacted BaCo$_2$(AsO$_4$)$_2$ powder. Details of the growth and x-ray diffraction are described in Ref. \onlinecite{zho20a}. Resulting crystals were dark pink plates of few mm across the samples and sub-mm thickness.

Since  BaCo$_2$(AsO$_4$)$_2$ is an insulator, we are not able to use electrical transport measurements under pressure. Instead, the measurements of two thermodynamic properties, magnetization and specific heat were employed to follow the changes in the ordering temperature and base temperature metamagnetism under pressure. 

Magnetization measurements at ambient and high pressure were performed for $H \| ab$ in a Quantum Design Magnetic Property Measurement System, MPMS 3, SQUID magnetometer. At ambient pressure the crystal was fixed on a quartz sample holder with a small amount of Dow Corning high vacuum grease. High pressure measurements up to $\sim 14$ kbar were performed in a commercial, HMD, Be-Cu, piston-cylinder cell \cite{hmd} with Daphne 7373 oil, which solidifies at $\sim 22$ kbar at room temperature \cite{yok07a}, as a pressure medium. The sample was oriented by eye with $H \| ab$ in the pressure cell. Magnetization data were taken upon increasing and decreasing pressure. No difference between the data taken upon increasing and decreasing pressure was found. This indicates that the orientation of the crystal did not change detectably through the pressure cycle.

Specific heat measurements under pressure up to $\sim 21$ kbar were performed using an AC calorimetry technique in a Quantum Design Physical Property Measurement System (PPMS).  Details of the setup used and the measurement protocol are described in Ref. \onlinecite{gat19a}. The sample was oriented as $H \| ab$. For magnetic field sweeps, a rate of 10 Oe/s was used. The measurements were performed in a  Be-Cu/Ni-Cr-Al hybrid piston-cylinder cell, similar to the one described in Ref. \onlinecite{bud86a}. A 40:60 mixture of light mineral oil:n-pentane, which solidifies at room temperature in the range of 30-40 kbar \cite{kim11a,tor15a}, was used as a pressure medium.

Given that (i) our ambient pressure measurements of $H1$ and $H2$ agreed well with literature and (ii) that our pressure dependent values of $H1$ and $H2$ for multiple measurements on multiple samples with arbitrary in-plane orientations of applied field also agreed well (see figure \ref{F7} below), we do not further consider possible in-plane anisotropy.

For both measurements the pressure values were determined by measuring superconducting transition of elemental Pb. \cite{eil81a}

Additionally, field dependent heat capacity at ambient pressure was measured utilizing the relaxation technique with fitting of the whole temperature response of the microcalorimeter, as implemented in the Heat Capacity option of a PPMS. In these measurements the magnetic field was stabilized at all points. To ensure the $H \| ab$ orientation, the sample was mounted on a small L-shaped platform made of oxygen-free copper foil. The contribution of the platform to the $C_p(H)$ data was negligible and had no effect on the signal associated with the metamagnetic transitions.

\section{Results}

Prior to a discussion of our pressure-dependent data sets, we first outline our choice of criteria to infer the ordering temperature $T_N$ from the temperature-dependent magnetization and specific heat data, as well as the ones to determine the critical fields $H1$ and $H2$ from the field-dependent data sets.

In general, the criteria for an antiferromagnetic, second order, transition from $M(T)$ and $C_p(T)$ were discussed in Ref. \onlinecite{fis62a}. Throughout this manuscript, for convenience, we will use the maxima in $d(MT)/dT$ and $C_p(T)$ to define the values of the N\'eel temperature $T_N$.
 
\begin{figure}
\begin{center}
\includegraphics[angle=0,width=120mm]{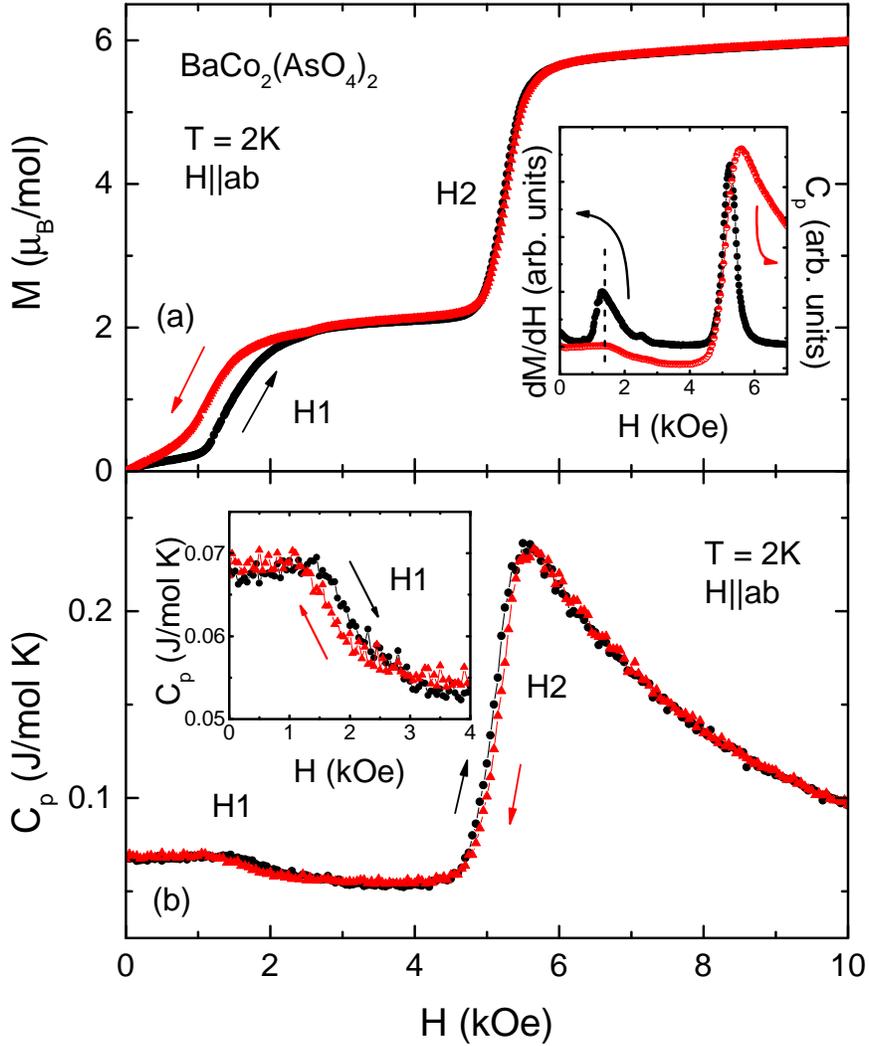}
\end{center}
\caption{(color online) (a) Field-dependent magnetization, $M(H)$ and (b) field-dependent specific heat, $C_p(H)$, of BaCo$_2$(AsO$_4$)$_2$  both measured at $T = 2$ K, $H \| ab$,  on increase and decrease of magnetic field. $H1$ and $H2$ mark lower and higher metamagnetic transitions respectively. Inset to (a): $dM/dH$ and $C_p$ measured on increase of magnetic field, plotted together for comparison. Inset to (b): low field part of the $C_p(H)$ data. All data are taken at ambient pressure.} \label{F1}
\end{figure}

Whereas the criteria used in temperature - dependent specific heat and magnetic susceptibility, as well as in field - dependent magnetization measurements are outlined in many publications and are reasonably well established, it is less so the case for the magnetic field - dependent specific heat measurements. To provide a consistent choice of criteria between the different data sets, we compare in Fig. \ref{F1} the ambient pressure, field-dependent magnetization and specific heat measured at $T = 2$\,K, with magnetic field applied in the $ab$ plane, and with the measurements done both on increase and decrease of magnetic field. The $M(H)$ data are consistent with the literature \cite{reg77a,zho20a}. The lower metamagnetic transition, $H1$, has a clear hysteresis. For $H1\,<H\,<H2$, the magnetization reaches $\sim 2\,\mu_B$/mol. The hysteresis of the second, final, metamagnetic transition is negligible and $M(H)$  above $H2$ has a value close to 6\,$\mu_B$/mol.  The specific heat data clearly show features associated with both metamagnetic transitions. The lower transition is seen as a shoulder. Specific heat in the antiferromagnetic phase has basically no field dependence. On further increase of magnetic field through the $H1$ transition, $C_p$ decreases.  In contrast, a rather large peak in the $C_p(H)$ corresponds to the upper $H2$ transition. Whereas in magnetization the position of the peak in $dM/dH$ is often taken as the criterion for the critical field of a metamagnetic transition, it appears that (at least in our case) the positions of the shoulder and the maximum in the $C_p(H)$ [see inset to Fig. \ref{F1}(a)] could approximate the lower and upper critical fields reasonably well. The slight difference that is apparent for the upper transition could be either specific for the measurements, or caused by slight misorientation of the samples with respect to the applied field in these measurements.

\begin{figure}
\begin{center}
\includegraphics[angle=0,width=120mm]{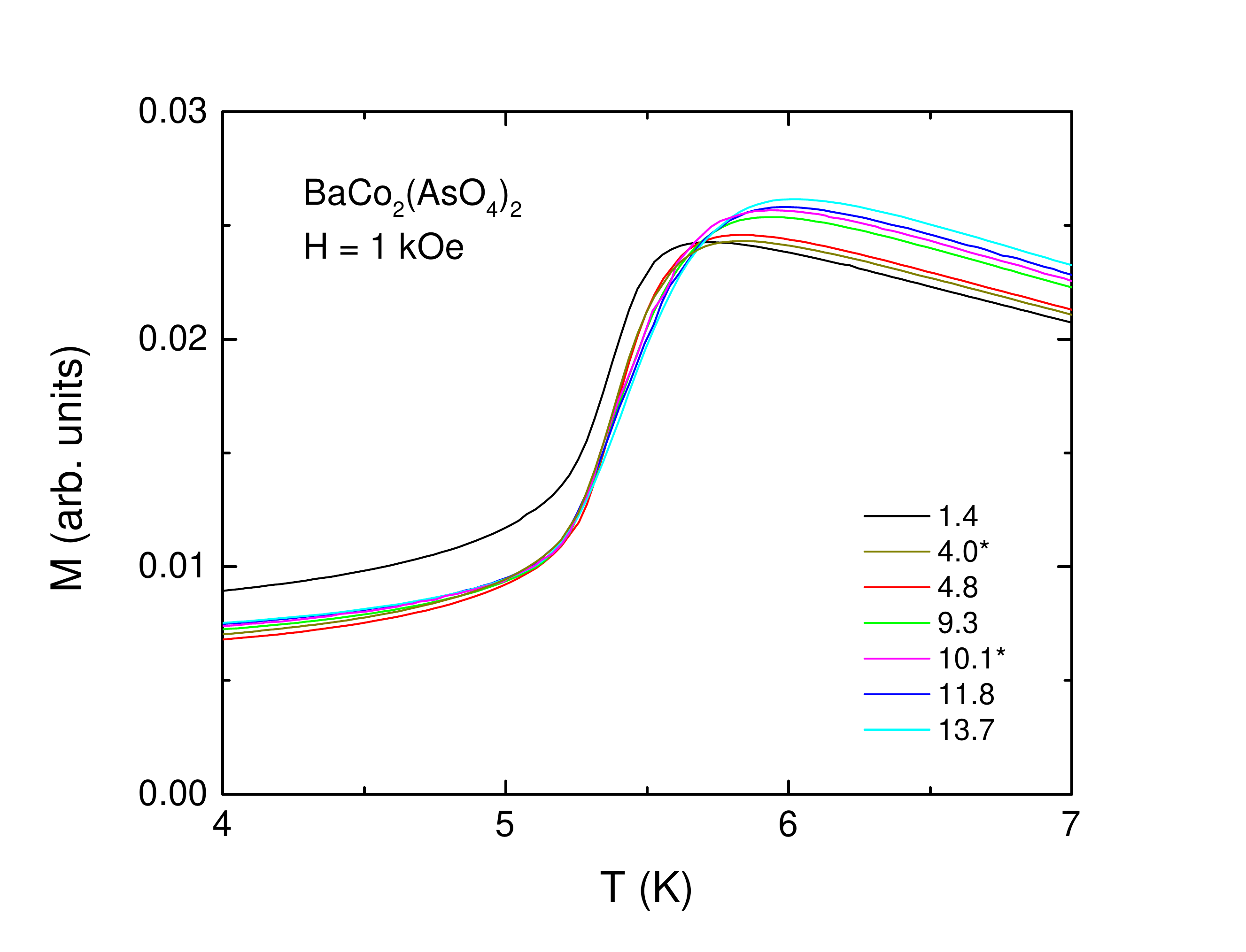}
\end{center}
\caption{(color online) Low temperature magnetization of  BaCo$_2$(AsO$_4$)$_2$ , measured at different pressures in the applied field of 1 kOe ($H \| ab$). The legend lists the pressure values in kbar. Asterisks points to measurements on pressure decrease.} \label{F2}
\end{figure}

\begin{figure}
\begin{center}
\includegraphics[angle=0,width=120mm]{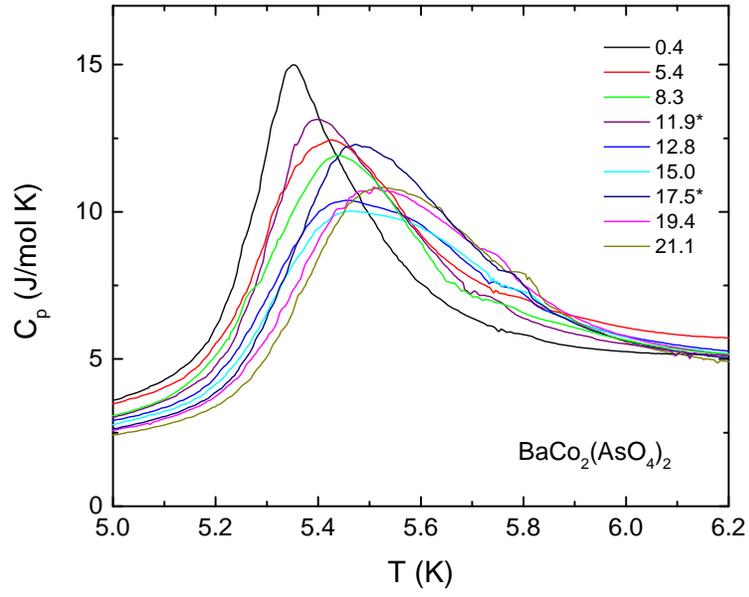}
\end{center}
\caption{(color online) Low temperature specific heat of  BaCo$_2$(AsO$_4$)$_2$ , measured at different pressures. The legend lists the pressure values in kbar. Asterisks points to measurements on pressure decrease.} \label{F3}
\end{figure}

Now we turn to our high-pressure results. Fig. \ref{F2} presents the temperature dependence of magnetization of  BaCo$_2$(AsO$_4$)$_2$ crystal below 10\,K and a field of 1\,kOe under different pressures up to $\sim 14$~kbar.  The magnetic ordering temperature increases under pressure, but only slightly. The $M(T)$ measurements in 20 Oe field (not shown here), show very similar increase of the N\'eel temperature as in 1 kOe. The specific heat measurements under pressure performed up to $\sim 21$~kbar (Fig. \ref{F3}) show a similar trend. 

\begin{figure}
\begin{center}
\includegraphics[angle=0,width=120mm]{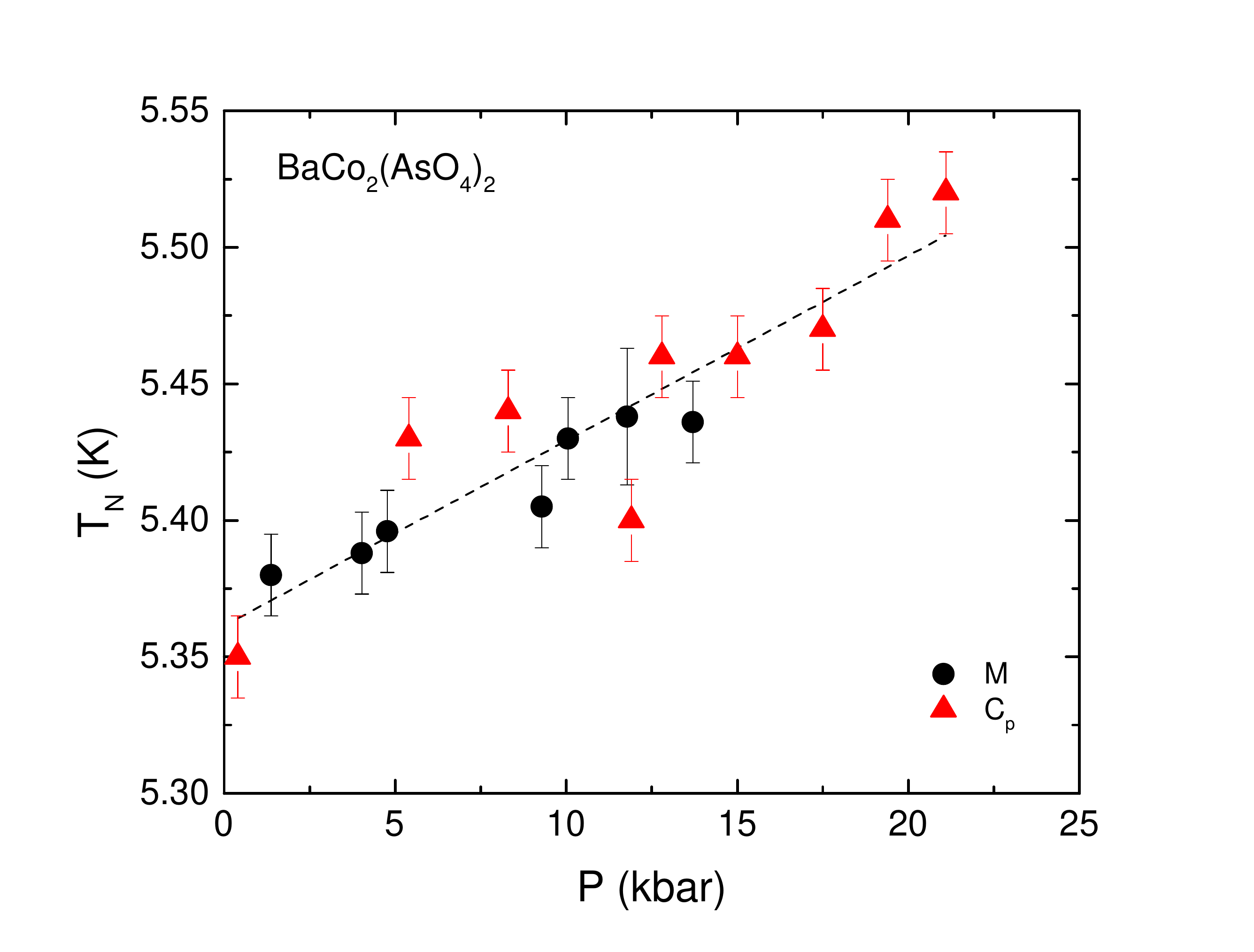}
\end{center}
\caption{(color online) Magnetic ordering temperature of  BaCo$_2$(AsO$_4$)$_2$ as a function of pressure. Circles - from maxima in $d(MT)/dT$, triangles - from maxima in $C_p(T)$. The error bars reflect temperature steps of the measurements and the level of signal to noise. Error bars in the pressure values do not exceed the size of the symbols.} \label{F4}
\end{figure}

The evolution of $T_N$ with pressure, inferred from both measurements, are combined in the phase diagram shown in Fig. \ref{F4}. The results of both measurements are consistent. The $T_N$ pressure derivative evaluated from the combination of both measurements is $dT_N/dP = 0.0068 \pm 0.0008$~K/kbar, so by 21 kbar the increase in $T_N$ is less than 3\%.

\begin{figure}
\begin{center}
\includegraphics[angle=0,width=120mm]{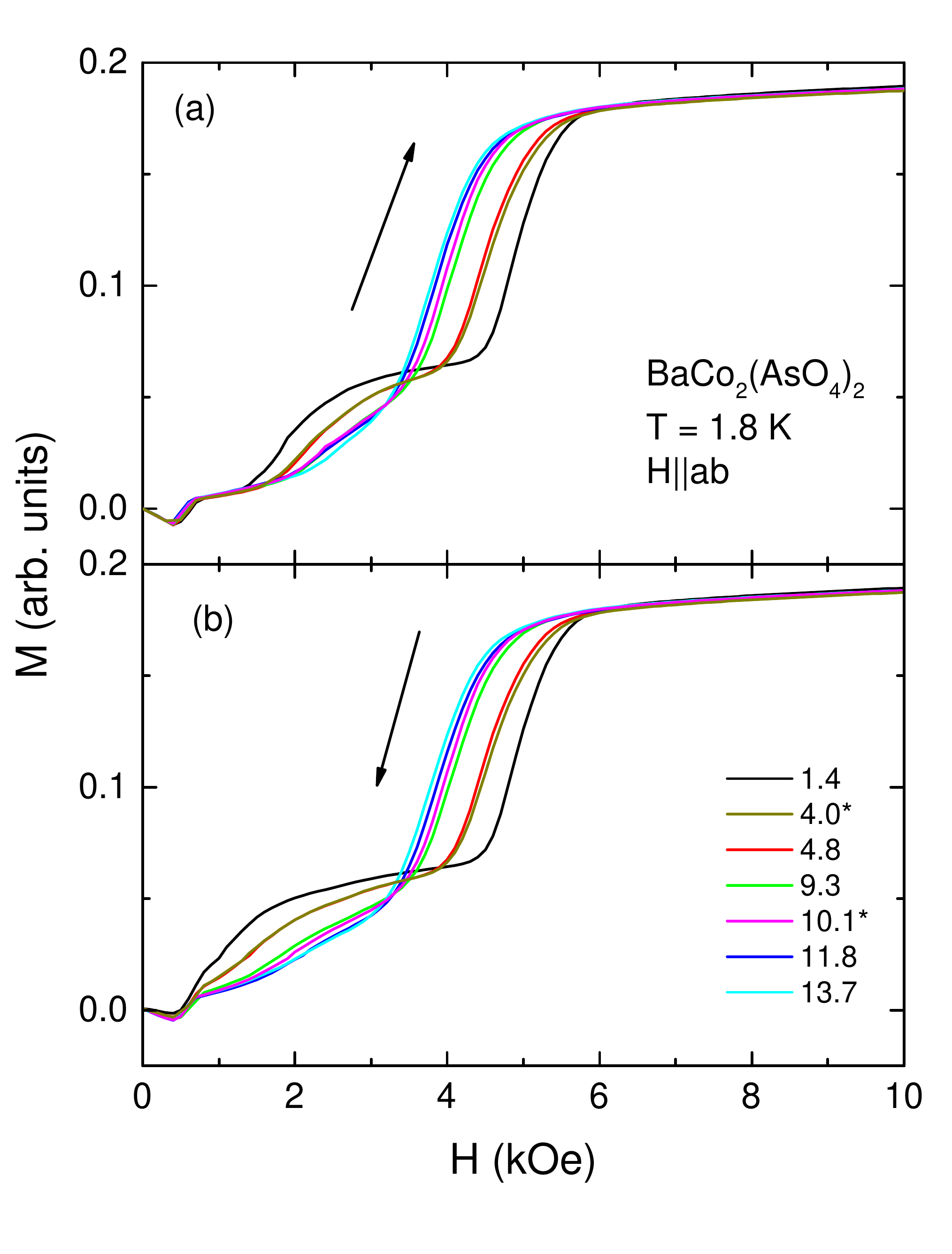}
\end{center}
\caption{(color online) Field-dependent, in-plane magnetization, $M(H)$, measured on field increase, (a), and field decrease, (b),  at $T = 1.8$~K at different pressures. The legend lists the pressure values in kbar. Asterisks points to measurements on pressure decrease.The small dips in $M(H)$ at low fields are due to the signal from elemental Pb.} \label{F5}
\end{figure}

\begin{figure}
\begin{center}
\includegraphics[angle=0,width=120mm]{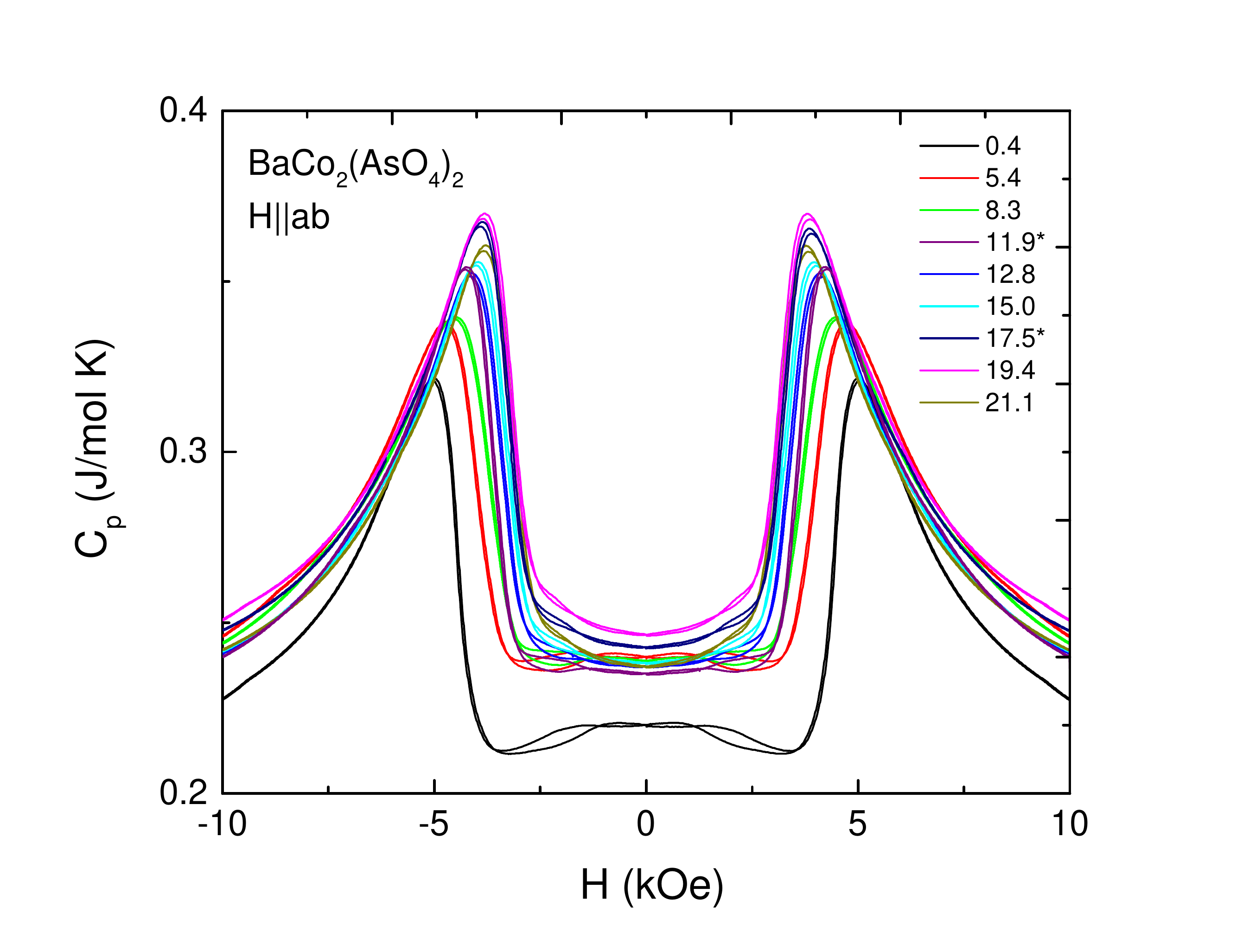}
\end{center}
\caption{(color online) Field-dependent specific heat, $C_p(H)$, measured in two quadrants on field increase and field decrease  at $T = 2$~K at different pressures. The legend lists the pressure values in kbar. Asterisks points to measurements on pressure decrease.} \label{F6}
\end{figure}

The effect of pressure on the critical fields is more significant. Figures \ref{F5} and \ref{F6} present field-dependent magnetization and specific heat, respectively. (Note that the overall change of the specific heat in magnetic field is qualitatively similar for the ambient pressure data measured using the relaxation technique and the AC calorimetry data under pressure. The difference in the $H = 0$ baseline could be due to details in sample mounts and implementation of these techniques.)  It is clear that under pressure the critical field of the  upper metamagnetic transition, $H2$, decreases, whereas the field $H1$ of the lower  transition increases. At the same time the lower metamagnetic transition becomes less hysteretic and its signatures in both measurements fade with the pressure increase. The resulting $P - H$ phase diagrams are summarized in Fig. \ref{F7}. For clarity and due to slightly different measurement temperatures as well as the different choices of criteria in $M(H)$ and $C_p(H)$ measurements (see Fig. \ref{F1}), two phase diagrams are plotted separately. Still they are telling the same story: by 15-20 kbar the value of $H2$ decreases by $\sim 25$ \%; $H1$ increases by $\sim 35$ \% on field-up and by $\sim 100$\% on field-down, the hysteresis in $H1$ gradually vanishes.

\section{Discussion and Summary}

To put the observed pressure dependencies into some context, we first review in more details the characteristics of the ambient-pressure magnetism of BaCo$_2$(AsO$_4$)$_2$ in zero and finite magnetic field, including its magnetoelastic effects, and derived proposed theoretical models.

BaCo$_2$(AsO$_4$)$_2$ is known to show strongly anisotropic magnetic behavior, with the in-plane susceptibiliy $\chi_{ab}$ greatly exceeding the out-of-plane susceptibility $\chi_{c}$. \cite{reg83a,reg90a,zho20a} In addition, it was commonly observed that the Curie-Weiss temperature is negative (positive) for fields applied out-of-plane (in-plane), indicating dominant antiferromagnetic (ferromagnetic) interactions in the respective direction. \cite{reg83a,reg90a,zho20a} Early elastic neutron scattering studies\cite{reg77a,reg90a} in zero field magnetic order suggested that BaCo$_2$(AsO$_4$)$_2$ shows helimagnetic order with an incommensurate wave vector ${\bf k} \simeq [0.27, 0, -4/3]$ and a phase angle $\phi \simeq 80 \pm 5^{\circ}$. Based on early classical mean-field models, assuming Heisenberg interactions only, it was argued that the helimagnetic order can be understood by taking at least three exchange couplings $J_1, J_2$ and $J_3$ into account with dominant ferromagnetic nearest-neighbor interaction and an antiferromagnetic interaction between more distant neighbors \cite{reg77a,reg90a,ras79a,mar12a}. Although somewhat different values of $J_1,  J_2, J_3$ are cited in the literature, e.g. 36 K, 4 K, -18 K (Ref. \onlinecite{reg77a}) or 38 K, 1.3 K, -10 K (Refs. \onlinecite{reg90a,mar12a}) (using the convention of $J > 0$ for ferromagnetic exchange interaction and $J < 0$ for antiferromagnetic), the general characteristic is that $J_1\,>\,0$ is dominant ($J_1\,\gg\,J_2\,>\,0$) and $J_3\,<\,0$ with $\lvert J_3/J_1 \rvert \sim 0.25-0.5$. That the properties of BaCo$_2$(AsO$_4$)$_2$ might indeed be described by a Heisenberg $XXZ$ model with significant frustration from third-neighbor antiferromagnetic interaction was also recently inferred from \textit{ab initio} calculations. \cite{das21a} It is noteworthy that in the classical and quantum model calculations \cite{fou01a} of the $J_1-J_2-J_3$ model, the proposed helimagnetic order in zero field only exists in a very narrow region of $J_2/J_1$ and $J_3/J_1$, which would imply that the ground state should be sensitive to external perturbations, if this model would accurately describe the properties of BaCo$_2$(AsO$_4$)$_2$. 

Elastic neutron measurements in finite field identified that the magnetic structure between $H1$ and $H2$ is characterized by a ferrimagnetic component and wavevectors ${\bf k_1} = [1/3, 0, -4/3]$ and ${\bf k_2} = 0$. \cite{reg79a,reg90a} This structure can be viewed as  a stacking of ferromagnetic chains parallel and antiparallel to the field in a + + -- + + -- ... sequence, with phases between two adjacent layers being random. \cite{reg90a}.  Assuming that the state above $H_2$ is corresponding to the saturated state with magnetization $m_{sat}$ (an assumption that is not consistent with the proposed Kitaev spin liquid), then the magnetization of the intermediate field phase is expected to be $m_{sat}$/3, in line with experimental observations. In the early papers (see e.g. Ref. \onlinecite{reg90a}) it was argued that the intermediate field phase can also be understood in terms of the $J_1-J_2-J_3$ model on a mean-field level.

Nonetheless, it must be said that not all aspects of the magnetism of BaCo$_2$(AsO$_4$)$_2$ could be successfully explained by the above-mentioned models. For example, the ordering temperature is much lower than the models would predict \cite{reg77a,reg90a} and also, the exchange couplings inferred from the \textit{ab initio} calculations are much larger than the ones found experimentally.\cite{das21a} Thus, in this regard, the pressure dependencies of $T_N$, $H1$ and $H2$ that we reveal here in this work represent another key benchmark for refining and testing of the theoretical models of the magnetic behavior of this compound.

They key result of our work of sizable changes of $H1$ and $H2$ with pressure is consistent with the idea that there are multiple, nearly degenerate, magnetic states present in BaCo$_2$(AsO$_4$)$_2$ with energy differences that are sensitive to small perturbations. What is remarkable is the fact that even though $T_N$ does not change significantly with pressure, the relative stability of the three phases observed in applied field do change significantly. Some insight into the microscopic effect of pressure can sometimes be inferred from considering the change of lattice parameters upon crossing the magnetic transition temperature $T_N$ at ambient pressure. Intriguingly, it was shown that the in-plane $a$ axis in  BaCo$_2$(AsO$_4$)$_2$ shows an anomalous negative thermal expansion in zero field \cite{mar12a,uwa21a}, i.e., the $a$ axis increases upon entering the helimagnetic magnetic state. This is in contrast to the behavior of the related compounds BaNi$_2$(AsO$_4$)$_2$ and  BaNi$_2$(PO$_4$)$_2$, where the $a$ axis decreases upon entering their respective magnetic ground states (Note that these two compounds show different magnetic structures at low temperatures). Taking this result at face value, this implies that the helimagnetic ground state, that might be stabilized by a sizable antiferromagnetic $J_3$ in the $J_1$-$J_2$-$J_3$ model, is favoured by a larger in-plane lattice parameter. Suppressing this distortion by the application of hydrostatic pressure is then naturally expected to weaken $J_3$ compared to $J_1$. As a result, within this picture, one would naively expect that hydrostatic pressure would move the system closer to the ferromagnetic/fully-saturated state. This is in line with our experimental observation of the decrease of $H2$ with pressure. 

However, it is then remarkable that $H1$ increases with increasing pressure. Given the narrow range of stability for helimagnetic order in the $J_1-J_2-J_3$-model together with the tentative explanation of decrease of $H2$ due to smaller $|J_3/J_1|$, this is counter intuitive at first. However, so far, in our discussion within this specific model, we ignored the impact of hydrostatic pressure on $J_2$, which is hard to estimate. The classical models certainly suggest that a decrease of $J_3$ can be compensated by a change in $J_2$. 

Note that these tentative explanations for our observed pressure-induced changes rely on simplified models that, e.g., omit non-Heisenberg interactions as well as interlayer couplings and assume no changes in $g$ factors. These ingredients are all important for a complete microscopic modeling of the pressure behavior. For example, explaining the marginal response of $T_N$ to pressure will certainly require an understanding of the three-dimensional coupling scheme.

Clearly, the next step would be to follow the evolution of $H1$ and $H2$ to higher pressures beyond 20\,kbar. Questions of interest, which would provide further benchmarks for microscopic modeling include, but are not limited to: (i) do higher pressures suppress the intermediate-field ferrimagnetic state completely?, (ii) can the zero-field helimagnetic order be suppressed by pressure and if so, (iii) what is the magnetic structure of BaCo$_2$(AsO$_4$)$_2$ then? We believe that our results encourage the performance of these measurements in the future. Ideally, these measurements would involve elastic and inelastic neutron scattering, and x-ray scattering under pressure, as well as thermodynamic and thermal transport (e.g. thermal Hall) measurements. It is also evident that further theoretical understanding of the microscopics of BaCo$_2$(AsO$_4$)$_2$  in terms of Heisenberg vs. non-Heisenberg interactions at ambient as well as finite pressure are needed.

\begin{acknowledgments}

We acknowledge useful discussions with S.M. Winter. Work at the Ames Laboratory was supported by the U.S. Department of Energy, Office of Science, Basic Energy Sciences, Materials Sciences and Engineering Division. The Ames Laboratory is operated for the U.S. Department of Energy by Iowa State University under contract No. DE-AC02-07CH11358. JS was supported in part by was supported in part by the Gordon and Betty Moore Foundation’s EPiQS Initiative through Grant No. GBMF-4411. EG gratefully acknowledges financial support from the Max Planck Society and the Deutsche Forschungsgemeinschaft through SFB 1143 (Project ID 247310070) through project C09. Work in Princeton (RZ and RJC) was supported by the Gordon and Betty Moore foundation Grant No GBMF-4412. 
SLB would like to acknowledge Russian Colloquium on Modern Problems of Condensed Matter Physics (\url{http://www.cond-mat.ru/}) for help in initiating this project.
\end{acknowledgments}

\clearpage

\clearpage

\clearpage

\clearpage

\clearpage

\clearpage

\clearpage

\clearpage

\begin{figure}
\begin{center}
\includegraphics[angle=0,width=120mm]{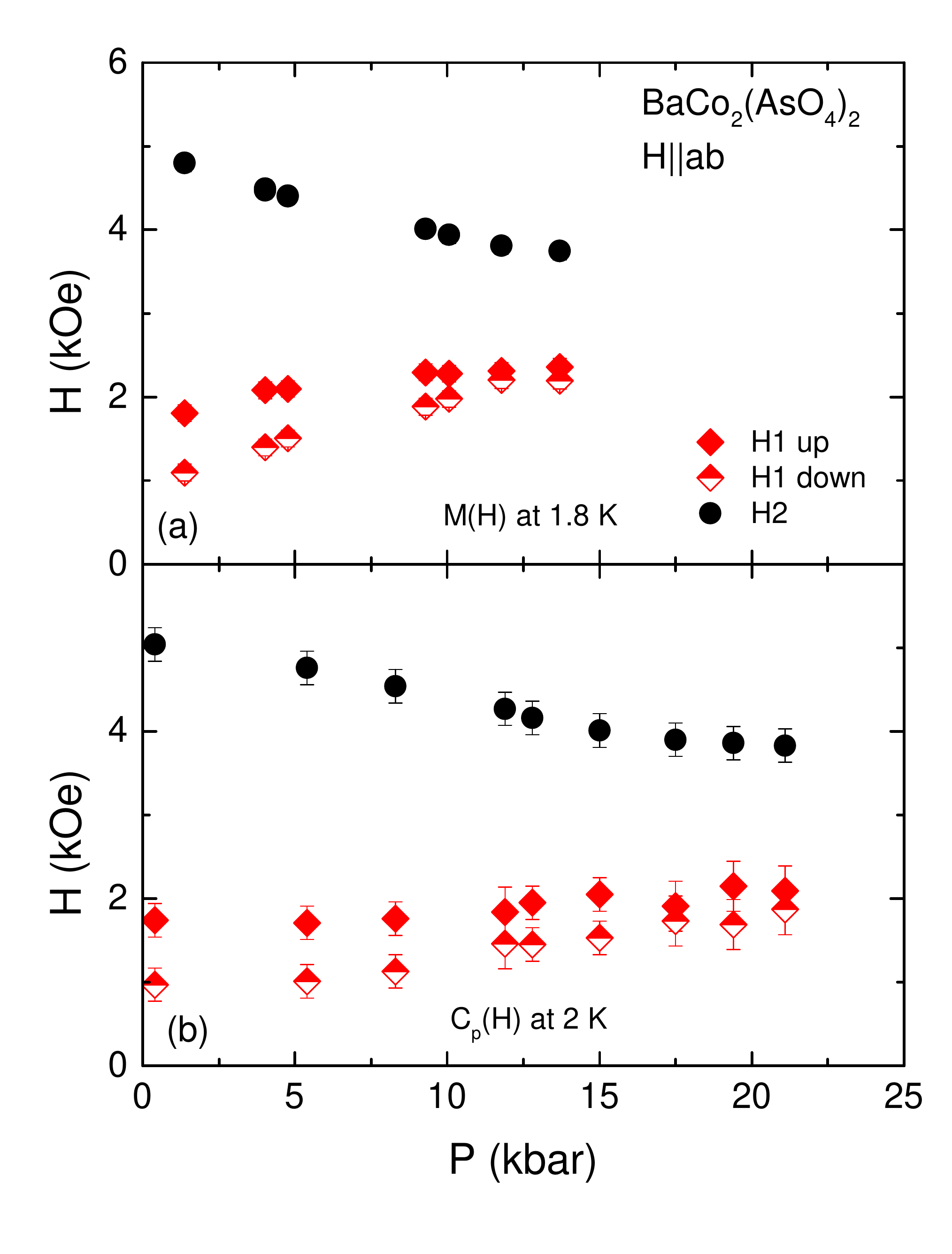}
\end{center}
\caption{(color online)Low temperature  $P - H$ phase diagram of  BaCo$_2$(AsO$_4$)$_2$: (a) from magnetization measurements; (b) from specific heat measurements. The lower metamagnetic transition $H1$ has hysteresis: both field up and field down points are plotted. The hysteresis of the upper transition is smaller than the size of the symbols.  Error bars in the pressure values do not exceed the size of the symbols.} \label{F7}
\end{figure}

\end{document}